\documentclass{jps-cp}
\usepackage{txfonts} 
\usepackage{bm}
\newcommand{\bea}{\begin{eqnarray}}
\newcommand{\eea}{\end{eqnarray}}

\newcommand{\nn}{\nonumber}

\title{Cubic, Non-Fermi Liquid Exchange Couplings\\for Doublet Impurities}

\author{Anna I. \textsc{T\'oth}, Davide \textsc{Marenduzzo}, Andrew D. \textsc{Huxley}}

\inst{Centre for Science at Extreme Conditions and SUPA, School of Physics and Astronomy,\\
University of Edinburgh, Edinburgh EH9 3FD, United Kingdom}

\email{atoth2{\_}ed.ac.uk}

\recdate{September 24, 2026}

\abst{We construct non-Fermi liquid (NFL) quantum impurity models for doublet impurities in a cubic metal, involving conduction electrons with half-integer total angular momentum, $J$.
  We highlight the only two crystal-symmetry-protected NFL quantum impurity models: the spin-half impurity, $J=3/2$ conduction electron Kondo model, and the two-channel Kondo model originally derived in the context of the quadrupolar Kondo model. We show explicitly that the quadrupolar Kondo model preserves time-reversal symmetry. We also construct two cubic quantum impurity models in which NFL behavior is not symmetry-protected. Each has $\Gamma_8$ and either  $\Gamma_{6}$ or  $\Gamma_{7}$ conduction electrons screening the impurity simultaneously, one model for a non-Kramers doublet and the other for a Kramers doublet impurity ground state. To our knowledge, the latter model has not been considered before; we expect it to realize a new NFL fixed point, but its solution is left for future work. Finally, we derive the relevant cubic perturbation that drives the non-symmetry-protected systems away from their NFL scaling regimes.  
}
\kword{Kondo model, non-Fermi liquid quantum impurities,  crystal field symmetry, time-reversal symmetry}

\begin{document}
\maketitle

\section{Introduction}
Since the 1980s, various forms of non-Fermi liquid (NFL) metallic behavior have been observed across a wide range of systems, including certain heavy fermion materials \cite{Ott83,Seaman91,Lohneysen94,Trovarelli00,Stewart,Nakatsuji08,Shimura15,Huang24}, high-$T_c$ cuprate \cite{Suzuki87,Gurvitch87} and iron-based \cite{Liu08,Kasahara10,Licciardello19} superconductors, engineered nanoelectronic quantum dot devices \cite{Potok07,Iftikhar15,Iftikhar18}, and twisted bilayer graphene \cite{Cao20}. Yet solvable microscopic models describing NFL phenomena remain scarce. One class of only a handful of exactly solvable microscopic models consists of frustrated quantum impurity models, including the overscreened multichannel Kondo models \cite{Nozieres80}. These models predate the observation of the corresponding NFL states, whose existence was first hypothesized by Nozières and Blandin in their landmark paper \cite{Nozieres80}. These NFL states are highly unstable against small perturbations, making their experimental observation particularly challenging. They have been observed in quantum dot devices \cite{Potok07,Iftikhar15,Iftikhar18}, whereas evidence for their existence in metals with magnetic impurities\textemdash the setting for which the conventional Kondo model was originally devised \cite{deHaas34,Kondo64}\textemdash has remained more elusive. To date, the most convincing evidence of NFL quantum impurity physics in metals comes from a 2018 study of the cubic cage compound  Y$_{1-x}$Pr$_x$Ir$_2$Zn$_{20}$ \cite{Yamane18,Onimaru16}\textemdash arriving only after decades of effort by many groups to experimentally realise the theoretical proposal from 1987 \cite{Cox87} that has long guided this search.

The idea is that in a cubic metal, a quantum impurity with a non-Kramers doublet ground state coupled to $\Gamma_8$ conduction electrons realizes two-channel Kondo (2CK) physics \cite{Nozieres80}\textemdash the most studied and best understood type of NFL quantum impurity physics\textemdash where the two-channels of spin-half electrons are furnished by the $\Gamma_8$ quartet, and channel symmetry is a consequence of cubic symmetry.
Thus, in this setup, the 2CK fixed point is protected by the cubic point group against channel anisotropy, which is a relevant perturbation. Its presence is usually the main obstacle to reaching the 2CK scaling regime.
In our recent catalog of crystal-symmetry-protected NFL Kondo-type exchange models \cite{Toth25}, we included the 2CK scenario based on Cox's construction \cite{Cox87}, subject to one caveat:  2CK physics is realized only in a certain coupling-parameter regime of the effective model, a spatially anisotropic 2CK model. Clearing up a slight controversy about the labeling of the $\Gamma_8$ conduction electrons, we show how time-reversal symmetry (TRS) is preserved  in this model in Sec.\ \ref{sec:2CK}.  Apart from the 2CK model, our catalog contains only one more crystal-symmetry-protected NFL Kondo-type exchange model with half-integer $J$ conduction electrons: the  spin-half impurity, $J=\frac32$ conduction electron Kondo model.  This model, introduced in Sec.\ \ref{sec:10CK}, does not suffer from the previous caveat and therefore appears to be a more promising and robust candidate for realization in dilute heavy-fermion materials. 

In the search for solvable NFL Hamiltonians, recent work \cite{Patri20b} identified another NFL universality class for a cubic system of an impurity with a non-Kramers doublet ground state screened simultaneously by $\Gamma_8$ and $\Gamma_6$ or $\Gamma_7$ conduction electrons whose fixed point is, however, not protected by crystal symmetry as the screening channels are not equivalent. In the vein of our previous analysis of crystal-symmetry-protected NFL quantum impurity models, in Sec.\ \ref{sec:Patri_cp}, we derive the counterpart to this model that involves the same type of conduction electrons but an impurity with a Kramers doublet ground state. To our knowledge, the latter model is a new addition to the  set of known NFL quantum impurity models. Finally, we list the relevant terms that drive the non-symmetry-protected models away from their NFL fixed points. 

\section{Cox's two-channel Kondo model and its invariance under time-reversal symmetry}\label{sec:2CK}
To construct a tractable effective model, originally drawn up to describe UBe$_{13}$ at low temperatures but whose applicability to that system remains debated, Cox assumed the impurity to be in a time-reversal-invariant 
$f^2$-state whose ground state transforms as the $\Gamma_3$ non-Kramers doublet irreducible representation (irrep) of the cubic point group $O$ \cite{Cox87}. We denote the non-Kramers doublet basis states by $\,|\Gamma_{3}^{\,3z^2-r^2}\rangle\,,\,|\Gamma_{3}^{\sqrt{3}(x^2-y^2)}\rangle\,$.
These impurity states hybridize with $l=3$ partial-wave conduction electron states expanded around the impurity.
Cox used the $J=5/2$ part of these states, omitting the $J\,=\,\frac 7 2$ part, to construct the $\Gamma_8$ quartet, arriving at a cubic exchange Hamiltonian that can be mapped onto the spatially anisotropic 2CK model. This choice is motivated by the expectation that for $l=3$, due to spin-orbit coupling, the $J=\frac 5 2$ states lie lower in energy than the $J=\frac 7 2$ ones.
The $\Gamma_8$ quartet is created by the cubic irreducible tensor operator \cite{Onodera66}
\bea\label{eq:Gamma_8_J_5/2}
{\bm \Psi^{\Gamma_8\,\dagger}_{J=\frac5 2}}\equiv\left(
\renewcommand{\arraystretch}{1.35}
\begin{array}{c}
\Psi_{\frac 3 2}^{\Gamma_8\,\dagger}\\
\Psi_{\frac 1 2}^{\Gamma_8\,\dagger}\\
\Psi_{-\frac 1 2}^{\Gamma_8\,\dagger}\\
\Psi_{-\frac 3 2}^{\Gamma_8\,\dagger}
\end{array}
\renewcommand{\arraystretch}{1}
\right)\equiv
\left(
\begin{array}{c}
-\,\frac 1{\sqrt 6}\,\psi_{\frac 3 2}^{\dagger}\,-\,\sqrt{\frac 5 6}\,\psi_{-\frac 5 2}^{\dagger}\\
\psi_{\frac 1 2}^{\dagger}\\
-\,\psi_{-\frac 1 2}^{\dagger}\\
\frac 1{\sqrt 6}\,\psi_{-\frac 3 2}^\dagger\,+\,\sqrt{\frac 5 6}\,\psi_{\frac 5 2}^\dagger
\end{array}
\right)
\eea
with $\psi_{J_z}^{\dagger}\,\equiv\,\psi_{J_z}^{J=\frac 5 2\,\dagger}$. This form does not depend on the value of $l$.
We use the following definition for the time-reversal operator:  $\,{\cal T}\,\equiv\,-\,i\sigma_y\,{\mathcal K}\,$,  where ${\mathcal K}$ is the complex conjugation and $\sigma_y$ acts only on the spin part. This way the different $\Gamma_8$ components are connected the same way as for a $J=3/2$ multiplet, a $\Gamma_8$ irrep in a cubic field, i.e.\ as
\bea\label{eq:TRS}
{\cal T}\psi_{J_z}^{J\,\dagger}{\cal T}^{-1}=(-1)^{J-J_z}\psi_{-J_z}^{J\,\dagger}\,. 
\eea
The adjoint quartet, ${\bm \Psi^{\Gamma_8}_{J=\frac5 2}}$, made up of
single-electron annihilation operators, can be defined as \cite{Messiah}
\bea
{\bm \Psi_{J=\frac5 2}^{\Gamma_8}}
\equiv\left(
\renewcommand{\arraystretch}{1.35}
\begin{array}{c}
\left({\Psi_{-\frac 3 2}^{\Gamma_8\,\dagger}}\right)^\dagger\\
-\left({\Psi_{-\frac 1 2}^{\Gamma_8\,\dagger}}\right)^\dagger\\
\left({\Psi_{\frac 1 2}^{\Gamma_8\,\dagger}}\right)^\dagger\\
-\left({\Psi_{\frac 3 2}^{\Gamma_8\,\dagger}}\right)^\dagger
\end{array}
\renewcommand{\arraystretch}{1}
\right)\equiv\left(
\renewcommand{\arraystretch}{1.35}
\begin{array}{c}
{\Psi_{-\frac 3 2}^{\Gamma_8}}\\
-\,{\Psi_{-\frac 1 2}^{\Gamma_8}}\\
{\Psi_{\frac 1 2}^{\Gamma_8}}\\
-\,{\Psi_{\frac 3 2}^{\Gamma_8}}
\end{array}
\renewcommand{\arraystretch}{1}
\right)=
\left(
\begin{array}{c}
\frac 1{\sqrt 6}\,\psi_{-\frac 3 2}^{}\,+\,\sqrt{\frac 5 6}\,\psi_{\frac 5 2}^{}\\
\psi_{-\frac 1 2}\\
\psi_{\frac 1 2}\\
\frac 1{\sqrt 6}\,\psi_{\frac 3 2}^{}\,+\,\sqrt{\frac 5 6}\,\psi_{-\frac 5 2}^{}
\end{array}
\right)\,.
\eea

Based on the following two cubic tensor product decompositions for the impurity and the conduction electron parts
\bea
\Gamma_3\,\otimes\,\Gamma_3&=&\Gamma_1\,\oplus\,\Gamma_2\,\oplus\,\Gamma_3\label{eq:G3xG3}\\
\Gamma_8\,\otimes\,\Gamma_8&=&\Gamma_1\,\oplus\,\Gamma_2\,\oplus\,\Gamma_3\,\oplus\,2\Gamma_4\,\oplus\,2\Gamma_5\,,\label{eq:G8xG8}
\eea
we can construct three independent, cubic-symmetry-allowed terms in the exchange Hamiltonian \cite{Toth25}.
The first term 
\bea
    {\cal H}^{\textrm{\,Cox}}_{\,\Gamma_1\otimes\,\Gamma_1}={\cal J}^{}_{\Gamma_1\otimes\,\Gamma_1}\sum_{\substack{\delta\in\left\{3z^2-r^2,\right.\\\left.\sqrt{3}(x^2-y^2)\right\}}}|\Gamma_{\,3}^{\,\delta}\rangle\langle\Gamma_{\,3}^{\,\delta}|\sum_{\mu=-3/2}^{3/2}n^{\,\Gamma_8}_{\,\mu}\,,
\eea
with $n^{\,\Gamma_8}_{\,\mu}\,\equiv\,\Psi_{\mu}^{\Gamma_8\,\dagger}\Psi^{\Gamma_8}_{\mu}$, describes potential scattering and is diagonal in the impurity and local conduction electron degrees of freedom. It is a marginal perturbation  
to the fixed points of the second and third cubic scalars that are 
\bea
{\cal H}^{\textrm{\,Cox}}_{\,\Gamma_2\otimes\,\Gamma_2}&=&{\cal J}_{\Gamma_2\otimes\,\Gamma_2}^{}\,\left(|\Gamma_{\,3}^{\,\alpha}\rangle\langle\Gamma_{\,3}^{\,\beta}|\,-\,
|\Gamma_{\,3}^{\,\beta}\rangle\langle\Gamma_{\,3}^{\,\alpha}|\right)\,\left(\Psi_{-\frac 3 2}^{\,\Gamma_8\,\dagger}\Psi^{\,\Gamma_8}_{\frac 1 2}\,-\,\Psi_{-\frac 1 2}^{\,\Gamma_8\,\dagger}\Psi^{\,\Gamma_8}_{\frac 3 2}\,-\,h.c.\right)\,,\\
{\cal H}^{\textrm{\,Cox}}_{\Gamma_3\otimes\,\Gamma_3}&=&{\cal J}_{\Gamma_3\otimes\,\Gamma_3}^{}\,\left[\left(|\Gamma_{3}^{\,\beta}\rangle\langle\Gamma_{3}^{\,\beta}|\,-\,
|\Gamma_{3}^{\,\alpha}\rangle\langle\Gamma_{3}^{\,\alpha}|\right)\,\left(n^{\,\Gamma_8}_{-\frac 3 2}\,-\,n^{\,\Gamma_8}_{-\frac 1 2}\,-\,n^{\,\Gamma_8}_{\frac 1 2}\,+\,n^{\,\Gamma_8}_{\frac 3 2}\right)\right.\,\nn\\
&&+\,\left(|\Gamma_{\,3}^{\,\alpha}\rangle\langle\Gamma_{\,3}^{\,\beta}|\,+\,
|\Gamma_{\,3}^{\,\beta}\rangle\langle\Gamma_{\,3}^{\,\alpha}|\right)\,\left.\left(\Psi_{-\frac 3 2}^{\,\Gamma_8\,\dagger}\Psi^{\,\Gamma_8}_{\frac 1 2}\,+\,\Psi_{-\frac 1 2}^{\,\Gamma_8\,\dagger}\Psi^{\,\Gamma_8}_{\frac 3 2}\,+\,h.c.\right)\right]\,
\eea
with state labels  $\alpha\equiv3z^2-r^2,\beta\equiv\sqrt{3}(x^2-y^2)$. With the following identifications, 
\bea
S^{\textrm{$\Gamma_3$\,imp}}_{+}&\equiv&|\Gamma_3^{\,\beta}\rangle\langle \Gamma_3^{\,\alpha}|\,,\quad S^{\textrm{$\Gamma_3$\,imp}}_-\,\equiv\,\left(S^{\textrm{$\Gamma_3$\,imp}}_+\right)^\dagger,\quad S^{\textrm{$\Gamma_3$\,imp}}_x\,\equiv\,\frac 1 2\left(S^{\textrm{$\Gamma_3$\,imp}}_{+}\,+\,S^{\textrm{$\Gamma_3$\,imp}}_{-}\right)\,,\\
S^{\textrm{$\Gamma_3$\,imp}}_{z}&\equiv&\frac 1 2\left(|\Gamma_3^{\,\beta}\rangle\langle \Gamma_3^{\,\beta}|\,-\,|\Gamma_3^{\,\alpha}\rangle\langle \Gamma_3^{\,\alpha}|\right)\,,\\
s^1_{+}&\equiv&\Psi_{\frac 3 2}^{\,\Gamma_8\,\dagger}\Psi^{\,\Gamma_8}_{-\frac 1 2}\,,\quad s^1_z\equiv\frac 1 2\left(n^{\,\Gamma_8}_{\frac 3 2}\,-\,n^{\,\Gamma_8}_{-\frac 1 2}\right)\,,
\quad s^1_-\equiv s^{1\,\dagger}_+\,,\\
s^2_+&\equiv&\Psi_{-\frac 3 2}^{\,\Gamma_8\,\dagger}\Psi^{\Gamma_8}_{\frac 1 2}\,,\quad s^2_z\equiv\frac 1 2\left(n^{\,\Gamma_8}_{-\frac 3 2}\,-\,n^{\,\Gamma_8}_{\frac 1 2}\right)\,,
\quad s^2_-\equiv s^{2\,\dagger}_+\,,
\eea
where the vector operators, ${\vec S}^{\textrm{$\Gamma_3$\,imp}},\,{\vec s^{\,1}},\,{\vec s^{\,2}}$ satisfy the canonical commutation relations, 
we can map ${\cal H}^{\textrm{\,Cox}}_{\,\Gamma_2\otimes\,\Gamma_2}+{\cal H}^{\textrm{\,Cox}}_{\,\Gamma_3\otimes\,\Gamma_3}$ onto the spatially anisotropic 2CK model
\bea
    {\cal H}^{\textrm{\,Cox}}_{\,\Gamma_2\otimes\,\Gamma_2}=4\,{\cal J}^{}_{\Gamma_2\otimes\,\Gamma_2}\,S^{\textrm{$\Gamma_3$\,imp}}_y\left(s_y^1\,+\,s_y^2\right)\,,\quad {\cal H}^{\textrm{\,Cox}}_{\,\Gamma_3\otimes\,\Gamma_3}=4\,{\cal J}^{}_{\Gamma_3\otimes\,\Gamma_3}\left[S^{\textrm{$\Gamma_3$\,imp}}_x\left(s_x^1\,+\,s_x^2\right)+S^{\textrm{$\Gamma_3$\,imp}}_z\left(s_z^1\,+\,s_z^2\right)\right],\nn\\
\eea
with the upper indices 1 and 2 distinguishing the two channels.  Note that the indices $x,y,z$ do not refer to actual spatial directions, they are only effective spatial indices. Cubic symmetry ensures equal couplings to the two channels. Spatial anisotropy is an irrelevant perturbation around the 2CK fixed point \cite{Emery92,Affleck92}. However, when ${\cal J}_{\Gamma_2\otimes\,\Gamma_2}\gg{\cal J}_{\Gamma_3\otimes\,\Gamma_3}$, the effective spatial anisotropy is no longer a perturbation, ${\cal H}^{\textrm{\,Cox}}_{\,\Gamma_2\otimes\,\Gamma_2}$ is the dominant term in the Hamiltonian and the system behaves as a Fermi liquid even at low temperatures. As the ratio of ${\cal J}_{\Gamma_2\otimes\,\Gamma_2}$ to ${\cal J}_{\Gamma_3\otimes\,\Gamma_3}$ is
not dictated by symmetry, the coupling of a $\Gamma_3$ impurity to local $\Gamma_8$ conduction electrons does not necessarily result in 2CK physics.

The TRS invariance of the exchange Hamiltonians can be established using Eqs.\ \eqref{eq:Gamma_8_J_5/2}-\eqref{eq:TRS}. Both the impurity parts and the conduction electron parts  of ${\cal H}^{\textrm{\,Cox}}_{\,\Gamma_i\otimes\,\Gamma_i}\,\,(i=1,2,3)$ are TRS even. Note that the relative signs and the correct labeling of
the $\Gamma_8$ irreducible tensor operator components are essential for constructing a cubic invariant out of them, and for determining its TRS properties; these were given in Ref.\ \cite{Onodera66}. Note also that the same construction and mapping to the 2CK model can be repeated for partial-wave states with different orbital angular momentum $l$ and total angular momentum $J$.

\section{The spin-half impurity $J=\frac 3 2$ conduction electron Kondo model}\label{sec:10CK}
Apart from the 2CK construction above, there is only one more crystal-symmetry-protected NFL quantum impurity model with half-integer $J$ conduction electrons: the spin-half impurity $J=\frac 3 2$ conduction electron Kondo model, which is the exchange coupling between a $\Gamma_6$ or $\Gamma_7$ impurity ground state and local $\Gamma_8$ conduction electrons \cite{Toth25}. The construction of the cubic exchange Hamiltonian proceeds as in the previous section, except that the impurity ground-state space is now a $\Gamma_6$ state (the construction is identical for a $\Gamma_7$ impurity), where the cubic tensor product decomposition for the bilinear operators is
\bea
\Gamma_6\,\otimes\,\Gamma_6&=&\Gamma_1\,\oplus\,\Gamma_4\,.\label{eq:G6xG6}
\eea
Comparing this with Eq.\ \eqref{eq:G8xG8}, there are three independent cubic exchange terms in the Hamiltonian: a marginal potential scattering and two NFL couplings. One of them is the spin-half impurity $J=\frac 3 2$ conduction electron Kondo model
 given by
\bea
{\cal  H}^{\textrm{\,Kondo}}_{S^{\textrm{imp}}=\frac12,\,J^{\,\textrm{c}}=\frac32}&=&{\cal J}{\vec S}^{\textrm{\,$\Gamma_6$\,imp}}\,{\vec J}^{\textrm{\,\,$\Gamma_8$\,c}}\,,
\eea
where ${\vec S}^{\textrm{\,$\Gamma_6$\,imp}}$ is the spin-half spin operator for the impurity, and  ${\vec J}^{\textrm{\,\,$\Gamma_8$\,c}}$ is the $J=\frac 3 2$ total angular momentum vector operator for the local conduction electrons. According to Refs.\ \cite{Fabrizio96,Sengupta96}, this model has identical properties in the spin sector to the ten-channel Kondo (10CK) model. This model is also TRS even as now both ${\vec S}^{\textrm{\,$\Gamma_6$\,imp}}$ and ${\vec J}^{\textrm{\,\,$\Gamma_8$\,c}}$ are TRS odd. The second symmetry-allowed NFL exchange coupling is also TRS even and it flows to the same fixed point, as was  demonstrated by solving it with the numerical renormalization group in Ref.\ \cite{Toth25}.

\section{Cubic, non-Fermi liquid quantum impurity models without crystal-symmetry protection}\label{sec:Patri_cp}
In Ref.\ \cite{Patri20b}, a cubic NFL model that is not symmetry-protected was constructed, with a $\Gamma_3$ non-Kramers doublet impurity ground state screened simultaneously by $\Gamma_8$ and $\Gamma_6$ conduction electrons.
If one considers only the $\Gamma_8$ electrons, the resulting effective model is the 2CK model as demonstrated in Sec.\ \ref{sec:2CK}. Screening a $\Gamma_3$ impurity with a single channel of $\Gamma_6$ conduction electrons does not by itself give NFL behavior, since $\Gamma_6$ has only twofold degeneracy; however, when it is coupled with $\Gamma_8$ conduction electrons a new type of NFL behavior is expected to emerge, provided the relevant perturbations that are not excluded by cubic symmetry (see Eq.\ \eqref{eq:relpert}) are sufficiently small. The relevant tensor product decomposition is
\bea
\Gamma_6\otimes\Gamma_8&=&\Gamma_3\oplus\Gamma_4\oplus\Gamma_5\,.\label{eq:G6xG8}
\eea
Comparing this with Eq.\ \eqref{eq:G3xG3}, we see there is one additional exchange coupling that can be present in a cubic Hamiltonian.
Denoting the $\Gamma_6$ irreducible tensor operator and its components as
\bea
{\bm \Psi^{\Gamma_6\,\dagger}}\equiv\
\left(\renewcommand{\arraystretch}{1.35}
\begin{array}{c}
  {\psi^{\Gamma_6}_\uparrow}^\dagger\\
  {\psi^{\Gamma_6}_\downarrow}^\dagger
\end{array}\right)\,,
\eea  
the NFL exchange coupling is given by
\bea
    {\cal H}_{\Gamma_{3-8-6}}^{\,}={\cal J}_{\Gamma_{3-8-6}}^{}\left[S_x^{\textrm{$\Gamma_3$\,imp}}\left({\psi^{\Gamma_6}_{\downarrow}}^\dagger{\Psi^{\Gamma_8}_{\frac 3 2}}^{}
      \,-\,{\psi^{\Gamma_6}_{\uparrow}}^\dagger{\Psi^{\Gamma_8}_{-\frac 3 2}}^{}\,+\,h.c.\right)+S_z^{\textrm{$\Gamma_3$\,imp}}\left({\psi^{\Gamma_6}_{\downarrow}}^\dagger{\Psi^{\Gamma_8}_{-\frac 1 2}}^{}
      \,-\,{\psi^{\Gamma_6}_{\uparrow}}^\dagger{\Psi^{\Gamma_8}_{\frac 1 2}}^{}\,+\,h.c.\right)\right]\,.
    \eea

    Repeating the construction with a $\Gamma_6$ (or $\Gamma_7$) Kramers doublet in place of the $\Gamma_3$ non-Kramers doublet as the impurity ground state yields a counterpart exchange coupling, which we expect to be NFL (cf.\  Eqs.\ \eqref{eq:G6xG6}, \eqref{eq:G6xG8}) 
\begin{multline}
    {\cal H}_{{\Gamma_{6-8-6}}}^{\,}={\cal J}^{}_{\Gamma_{6-8-6}}\left[S_x^{\textrm{$\Gamma_6$\,imp}}\frac12\left(-\,{\psi^{\Gamma_6}_{\downarrow}}^\dagger{\Psi^{\Gamma_8}_{\frac 1 2}}^{}
      +\sqrt{3}\,{\psi^{\Gamma_6}_{\downarrow}}^\dagger{\Psi^{\Gamma_8}_{-\frac 3 2}}^{}-\sqrt{3}\,{\psi^{\Gamma_6}_{\uparrow}}^\dagger{\Psi^{\Gamma_8}_{\frac 3 2}}^{}\,+\,{\psi^{\Gamma_6}_{\uparrow}}^\dagger{\Psi^{\Gamma_8}_{-\frac 1 2}}^{}\,+\,h.c.\right)\right.\\
     -\,S_y^{\textrm{$\Gamma_6$\,imp}}\frac{i}{2}\left({\psi^{\Gamma_6}_{\downarrow}}^\dagger{\Psi^{\Gamma_8}_{\frac 1 2}}^{}
      +\sqrt{3}\,{\psi^{\Gamma_6}_{\downarrow}}^\dagger{\Psi^{\Gamma_8}_{-\frac 3 2}}^{}+\sqrt{3}\,{\psi^{\Gamma_6}_{\uparrow}}^\dagger{\Psi^{\Gamma_8}_{\frac 3 2}}^{}\,+\,{\psi^{\Gamma_6}_{\uparrow}}^\dagger{\Psi^{\Gamma_8}_{-\frac 1 2}}^{}\,-\,h.c.\right)\\
      \left.+\,S_z^{\textrm{$\Gamma_6$\,imp}}\left({\psi^{\Gamma_6}_{\downarrow}}^\dagger{\Psi^{\Gamma_8}_{-\frac 1 2}}^{}
      \,+\,{\psi^{\Gamma_6}_{\uparrow}}^\dagger{\Psi^{\Gamma_8}_{\frac 1 2}}^{}\,+\,h.c.\right)\right]\,.
    \end{multline}
Note that ${\cal H}_{{\Gamma_{6-8-6}}}^{\,}$ is already spherically symmetric, not merely cubic-symmetric. Its full SU(2) angular momentum symmetry makes its solution by the numerical renormalization group \cite{Wilson75} or a conformal field theory analysis more tractable, which we leave for future work. 

When the following two cubic, potential scattering terms are present in the Hamiltonian with couplings that are not equal, ${\cal J}_{\,\Gamma_1\otimes\Gamma_1}^{\,\Gamma_6\textrm{\,c}}\neq {\cal J}_{\,\Gamma_1\otimes\Gamma_1}^{\,\Gamma_8\textrm{\,c}}$, they break channel symmetry and drive the system away from the respective NFL fixed points of ${\cal H}_{{\Gamma_{3-8-6}}}^{\,}$ and ${\cal H}_{{\Gamma_{6-8-6}}}^{\,}$\,,
\bea
{\cal H}_{\,\Gamma_1\otimes\Gamma_1}^{\,\Gamma_6\textrm{\,c}}={\cal J}_{\,\Gamma_1\otimes\Gamma_1}^{\,\Gamma_6\textrm{\,c}}\left({\psi^{\Gamma_6}_\uparrow}^\dagger\,{\psi^{\Gamma_6}_\uparrow}^{}\,+\,{\psi^{\Gamma_6}_\downarrow}^\dagger\,{\psi^{\Gamma_6}_\downarrow}^{}\right)\,,\quad {\cal H}_{\,\Gamma_1\otimes\Gamma_1}^{\,\Gamma_8\textrm{\,c}}={\cal J}_{\,\Gamma_1\otimes\Gamma_1}^{\,\Gamma_8\textrm{\,c}}\sum_{\alpha=-\frac{3}{2}}^{\frac{3}{2}}\,{\Psi^{\Gamma_8}_{\alpha}}^\dagger{\Psi^{\Gamma_8}_{\alpha}}^{}\,.\label{eq:relpert}
\eea

\section{Conclusions}
We have considered cubic exchange models in which a doublet impurity couples to conduction electrons with half-integer total angular momentum $J$. Using the method of Ref.\ \cite{Toth25}, we have constructed four models with NFL fixed points at $T = 0$, the last of which is conjectured, and examined their robustness against cubic perturbations.  By giving the $\Gamma_8$ basis states and the exchange Hamiltonians explicitly, with the correct labeling and relative signs, we showed that Cox's quadrupolar Kondo model is time-reversal invariant. 

First, Cox's model (Sec.\ \ref{sec:2CK}) maps onto a spatially anisotropic 2CK model.  Cubic symmetry guarantees equal couplings to the two channels, so channel anisotropy, the main obstacle to 2CK physics, is excluded.  Spatial anisotropy is irrelevant at the 2CK fixed point, but when it is large enough it can still obscure the 2CK scaling regime and complicate experimental verification.  Second, the spin-half impurity, $J=\frac32$ conduction electron Kondo model (Sec.\ \ref{sec:10CK}) is also protected by crystal symmetry, and its spin sector is identical to that of the 10CK model.  Since it avoids the caveat of the first model, it has the greatest chance of the four of being realized experimentally.  Third, the two models of Sec.\ \ref{sec:Patri_cp}, namely the $\Gamma_3$ model of Ref.\ \cite{Patri20b} and the new Kramers-doublet counterpart derived here, whose NFL fixed point we expect but have not yet established, are allowed by cubic symmetry but are fragile: the difference between the cubic-symmetry-allowed potential scattering terms of Eq.\ \eqref{eq:relpert} will drive the system away from the respective NFL fixed points. The NFL fixed point of the new model is expected by analogy with the $\Gamma_3$ model but remains to be established, and its solution is left for future work.

\section{Acknowledgments}
We would like to thank the Referee for their helpful comments and for directing us to Ref.\ \cite{Onodera66}.
This project has received funding from the European Union’s Horizon 2020 research and innovation programme under the Marie Skłodowska-Curie grant agreement No 101024548.

\end{document}